\documentclass[12pt,oneside]{article}
\usepackage{amsfonts}
\usepackage{eufrak}
\usepackage[T1]{fontenc}
\usepackage[dvips]{graphicx}
\usepackage{float}
\title{Numerical Evaluation of some Parameters for a Model of Neutral Kaons}
\author{J.Jankiewicz\footnote{e--mail:
jjank@proron.if.uz.zgora.pl}\\ University of Zielona G\'{o}ra,
Institute of Physics\\ ul. Prof. Z. Szafrana 4a, 65--516 Zielona
G\'{o}ra, Poland}
\begin{document}
\bibliographystyle{plain}
\maketitle
\begin{abstract}
Using the Mathematica program we calculate numerically the
difference of the diagonal matrix elements of the time dependent
effective Hamiltonian for the neutral K meson complex. We consider
the exactly solvable neutral K meson model based on the one-pole
approximation for the mass density. The so-called Khalfin's Theorem
is numerically examined. Some characteristic parameters for this
system are also calculated. The results of all calculations are
presented in the graphical form. The calculations are made assuming
the total system is CPT -- invariant and CP -- noninvariant.
\end{abstract}

\section{Introduction}

The neutral kaon system is probably one of the most interesting
complexes of elementary particles.  Using this system it was found
in 1964 that the CP symmetry is violated \cite{Christensen}. The
description of CP violation effects in this system is based on the
approximation proposed by Lee, Oehme and Yang (LOY) in \cite{Lee}.
This theory was then developed and intensively studied by Lee (see
\cite{LOY}) and by many other authors (see eg. \cite{many}). Within
the LOY approach, a non--hermitian Hamiltonian $H_{\parallel} $ is
used to study the properties of the particle-antiparticle unstable
system \cite{Lee,LOY,many,p:31}
\begin{equation}
 H_{\parallel}\equiv M-\frac{i}{2}\Gamma, \label{j1-5}
\end{equation}
where
\begin{equation}
M=M^{+} \ , \  \Gamma = \Gamma^{+}  \label{j1-6}
\end{equation}
are $(2\times2) $ matrices acting in $\mathcal{H}_{\parallel}$,
where $\mathcal{H}_{\parallel}$ is a two--dimensional subspace of
the total Hilbert space of states $\mathcal{H}$, spanned by the
state vectors of $K^{0},$ $\bar{K}^{0} $ mesons. The $M $-matrix is
called the mass matrix and $\Gamma$ is the decay matrix. LOY derived
their approximate effective Hamiltonian $H_{\parallel}\equiv H_{LOY}
$ by adapting the one-dimensional Weisskopf-Wigner (WW) method to
the two-dimensional case corresponding to the neutral kaon system.
Almost all properties of this system can be described by solving the
Schr\"{o}dinger--like equation
\begin{equation}
i\frac{\partial }{\partial t}|\psi
;t\rangle_{\parallel}=H_{\parallel }|\psi ;t\rangle_{\parallel}, \ \
\  (t\geq t_{0}> -\infty ) \label{j1-7}
\end{equation}
where we have used  $\hbar =c=1 $, and $|\psi ;t\rangle_{\parallel}
\in \mathcal{H}_{\parallel}$.

Within the LOY theory the physical states of neutral kaons are
superpositions of $|K^{0}\rangle$ and $|\bar{K}^{0}\rangle$. They
are the eigenvectors of $H_{LOY}$,
\begin{eqnarray}
|K_{S}\rangle =p|K^{0}\rangle + q|\bar{K}^{0}\rangle, \ \
|K_{L}\rangle =p|K^{0}\rangle -q|\bar{K}^{0}\rangle , \label{j1-47}
\end{eqnarray}
\begin{equation}
H_{\parallel}|K_{S(L)}\rangle = \mu_{S(L)}|K_{S(L)}\rangle,
\label{mu-L-S}
\end{equation}
and correspond to short-living (the vector $|K_{S}\rangle$) and
long-living (the vector $|K_{L}\rangle$) states of neutral kaons. We
will use the following notations further on in this paper:
$|K^{0}\rangle \equiv |\textbf{1}\rangle $, $|\bar{K}^{0}\rangle
\equiv |\textbf{2}\rangle$.

One of standard results of the LOY approach is the following: In a
CPT invariant system, i.e. when
\begin{equation}
\Theta H \Theta^{-1}=H, \label{j1-3}
\end{equation}
(where $\Theta = CPT$, and $H$ is the total self-adjoint Hamiltonian
for the system containing neutral koans considered), there is
\begin{equation}
h_{11}^{LOY}=h_{22}^{LOY} \label{j1-14}
\end{equation}
and
\begin{equation}
M_{11}^{LOY}=M_{22}^{LOY}, \label{j1-15}
\end{equation}
where: $M_{jj}^{LOY}=\Re (h_{jj}^{LOY}) $ and $\Re (z)$ denotes the
real part of a complex number $z$ ($\Im (z) $ is the imaginary part
of $z$), and $h_{jj}^{LOY}=\langle \textbf{j}|H_{LOY} |\textbf{j}
\rangle $, $(j=1,2) $.  Another important prediction of the LOY
theory is that the ratio
\begin{equation}
r(t) \stackrel{\rm def}{=} \frac{p^{\,2}}{q^{\,2}} \equiv
\frac{A_{12}(t)}{A_{21}(t)} = {\rm const}\;\;\;{\rm and}\;\;\;
|r(t)|=|\frac{p^{\,2}}{q^{\,2}}| \neq 1, \label{r(t)}
\end{equation}
when CP symmetry is violated, $[CP, H] \neq 0$, \cite{LOY} --
\cite{khalfin-fp}. Here
\begin{equation}
A_{jk}(t)=\langle \textbf{j}|U_{\parallel}(t)|\textbf{k}\rangle
\equiv \langle
\textbf{j}|e^{\textstyle{-itH_{||}}}|\textbf{k}\rangle \ \
(j,k=1,2), \label{j1-23}
\end{equation}
and $U_{\parallel}(t)$ is the evolution operator for the subspace
${\cal H}_{\parallel}$.

The important result indicating some limitations of the LOY approach
was obtained by Khalfin
\cite{chiu,khalfin,khalfin-fp,kabir,kabir+mitra,dass,p:2}. Khalfin
found that in the exact theory there must be
\begin{equation}
{\rm if}\;\;\; r(t) = \frac{A_{12}(t)}{A_{21}(t)} = {\rm
const.},\;\;\;\;\;{\rm then}\;\;\; \;|r(t)| = 1, \label{khalfin}
\end{equation}
where
\begin{equation}
A_{jk}(t)= \langle
\textbf{j}|e^{\textstyle{-itH}}|\textbf{k}\rangle,\ \ \ (j,k=1,2).
\label{A-jk-exact}
\end{equation}
Result (\ref{khalfin}) is known in the literature as "Khalfin's
Theorem". Using this result Khalfin hypothesized that beyond the LOY
approximation one should expect new CP-violation effects
\cite{khalfin,khalfin-fp} and he tried to obtain some model
estimations of the possible magnitude of these effects. He found
that the order of these effects should be $10^{-3}$ (see
\cite{khalfin-fp}). He obtained his estimation using the spectral
language for the description of $K_{S}, K_{L}$ and $K^{0},$
$\bar{K}^{0} $, by introducing a hermitian Hamiltonian, $H$, with a
continuous spectrum of decay products labeled by $\alpha, \beta $,
etc.,
\begin{eqnarray}
H|\phi_{\alpha}(m)\rangle =m \, |\phi_{\alpha}(m)\rangle ,
\;\;\;\;\langle \phi_{\beta}(m')|\phi_{\alpha}(m) \rangle =
\delta_{\alpha \beta }\delta (m'-m). \label{j1-65}
\end{eqnarray}
Here $H$ is the above mentioned total Hamiltonian for the system.
$H$ includes all interactions and has absolutely continuous
spectrum. We have
\begin{eqnarray}
|K_{S}\rangle =\int_{{\rm Spec}\; (H)}dm \;
\sum_{\alpha}\omega_{S,\alpha}(m)|\phi_{\alpha}(m)\rangle ,
\label{j1-66}
\end{eqnarray}
\begin{eqnarray}
|K_{L}\rangle =\int_{{\rm Spec}\;(H)}dm \;
\sum_{\beta}\omega_{L,\alpha}(m)|\phi_{\beta}(m)\rangle,
\label{j1-67}
\end{eqnarray}
and
\begin{equation}
|\textbf{j}\rangle =\int_{{\rm Spec}\; (H)}dm \;
\sum_{\alpha}\omega_{j,\alpha}(m)|\phi_{\alpha}(m)\rangle ,
\label{Ko-bar-Ko}
\end{equation}
where $j= 1,2$. Thus, the exact $A_{jk}(t)$ can be written as the
Fourier transform of the density $\rho_{jk}(m),$ ($j,k=1,2$),
\begin{equation}
A_{jk}(t)= \int_{-\infty}^{+\infty} dm \; e^{-imt}\rho_{jk} (m),
\label{j1-37ab}
\end{equation}
where
\begin{equation}
\rho_{jk}(m) =
\sum_{\alpha}\omega_{j,\alpha}^{\ast}(m)\,\omega_{k,\alpha}(m).
\label{rho-jk}
\end{equation}
The minimal mathematical requirement for $\rho_{jk}(m)$ is the
following: \linebreak $ \int_{-\infty}^{+\infty} dm \,|\rho_{jk}(m)|
\;<\;\infty$. Other requirements for $\rho_{jk}(m)$ are determined
by basic physical properties of the system. The main property is
that the energy (i.e. the spectrum of $H$) should be bounded from
below, ${\rm Spec} (H) = [m_{g}, \,\infty)$ and $m_{g}
> - \infty$.

Starting from  densities $\rho_{jk}(m)$ one can calculate
$A_{jk}(t)$. In order to find these densities from relation
(\ref{rho-jk}) one should know the expansion coefficients
$\omega_{j,\alpha}(m)$. Using physical states $|K_{S}\rangle,
|K_{L}\rangle$ and relations (\ref{j1-47}) they can be expressed in
terms of the expansion coefficients $\omega_{S,\alpha}(m),
\omega_{S,\alpha}(m)$. Thus, assuming the form of coefficients
$\omega_{S,\alpha}(m), \omega_{S,\alpha}(m)$ defining physical
states of neutral kaons one can compute all $A_{jk}(t)$, ($j,k =
1,2$).

The model considered by Khalfin is based on the assumption that
(see formula (35) in \cite{khalfin-fp}),
\begin{equation}
\omega_{S,\beta}(m)= \sqrt{\frac{\Gamma_{S}}{2\pi}}\  \,
\frac{\xi_{S,\beta}(m)}{|\xi_{S,\beta}(m_{S} -
i\frac{\Gamma_{S}}{2})|}\,
\frac{a_{S,\beta}(K_{S}\rightarrow\beta)}{m-m_{S}+
i\frac{\Gamma_{S}}{2}}, \label{j1-72}
\end{equation}
\begin{equation}
\omega_{L,\beta }(m)=\sqrt{\frac{\Gamma_{L}}{2\pi}}\,
\frac{\xi_{L,\beta}(m)}{|\xi_{S,\beta}(m_{L} -
i\frac{\Gamma_{L}}{2})|}\,
\frac{a_{L,\beta}(K_{L}\rightarrow\beta)}{m-m_{L}+
i\frac{\Gamma_{L}}{2}}, \label{j1-73}
\end{equation}
where  $a_{S,\beta} $ and  $a_{L,\beta} $ are the decay (transition)
amplitudes and $\xi_{S(L),\beta}(m)$ are, in general, some
nonsingular "preparation functions". Khalfin found his above
mentioned estimation choosing, for simplicity, the trivial form of
the "preparation functions", $\xi_{S(L),\beta}(m) = 1$.

The discussion about the validity of the Khalfin's estimation of his
new CP violation effect can be found in the literature (see eg.
\cite{chiu,p:2}). Our attention will be concentrated on the attempt
to verify the size of the Khalfin's estimation performed in
\cite{p:2}. The calculation performed in \cite{p:2} uses Khalfin's
assumption that $\xi_{S(L),\beta}(m) = 1$, strictly speaking, they
use the assumption that  in (\ref{j1-72}), (\ref{j1-73}) there is
\begin{eqnarray}
\frac{\xi_{S(L),\beta}(m)}{|\xi_{S(L),\beta}(m_{S(L)} -
i\frac{\Gamma_{S(L)}}{2})|} &\equiv&  g(m - m_{g}) = [g(m -
m_{g})]^{2} \nonumber\\
&\stackrel{\rm def}{=}& \left\{
                      \begin{array}{cc}
                        1 & {\rm if}\; m \geq m_{g}, \\
                        0 & {\rm if}\; m  < m_{g},\\
                      \end{array}
                    \right. .\label{g(m)}
\end{eqnarray}
Within this assumption one obtains, for example, that
\begin{equation}
{\cal A}_{SS}(t)\stackrel{\rm def}{=}  \langle
K_{S}|e^{\textstyle{-itH}}|K_{S}\rangle =\int_{-\infty}^{+\infty}\;
dm \; \rho_{SS}(m)\,e^{\textstyle{-itm}} , \label{A-SS}
\end{equation}
where
\begin{equation}
\rho_{SS}(m)\,=\, g(m - m_{g})\, \frac{\Gamma_{S}}{(m-m_{S})^{2}+
\frac{\Gamma_{S}^{2}}{4}}\, \frac{S}{2\pi},\label{rho-SS}
\end{equation}
\begin{equation}
S = \sum_{\alpha}|a_{S,\alpha}(K_{S}\rightarrow\alpha)|^{2},
\label{s}
\end{equation}
and so on.

The form of  density $\rho_{SS}(m)$ defined by (\ref{rho-SS}) is not
the most general one. In more realistic models functions
$\omega_{j,\alpha}(m)$ and of type $\omega_{S,\beta}(m), \omega_{L,
\beta}(m)$ lead to the densities $\rho(m)$ which general form is
similar to (\ref{rho-SS}) with $g(m - m_{g})$ and $S \equiv S(m)$
having more involved form \cite{fonda,nowakowski2}. In the general
case the threshold factor $g(m - m_{g})$ describes the behavior of
$\rho (m)$ for small $m$, (i.e., for $m \simeq m_{g}$) and it is
responsible for the long time properties of amplitudes of type
$A_{jk}(t)$ and ${\cal A}_{SS}(t)$. The second factor in the
formulae of type (\ref{rho-SS}) having the Breit--Wigner form
results from the pole structure of functions of type
$\omega_{S,\beta}(m), \omega_{L,\beta}(m)$ (see (\ref{j1-72}),
(\ref{j1-73})) defining densities $\rho(m)$ and it is responsible
for the form of $A_{jk}(t), {\cal A}_{SS}(t)$ etc. for the
intermediate times (i.e. it is responsible for the exponential part
of the survival probabilities). The third factor, i.e. the factor
corresponding to $S(m)$ ensures the suitable behavior of $\rho (m)$
for $m \rightarrow \infty$.

For simplicity, it is assumed in \cite{p:2} that $m_{g} = 0$. So all
integrals of type (\ref{A-SS}) and (\ref{j1-37ab}) are taken between
the limits $m=0$ and $m= + \infty$. All these assumptions made it
possible to express amplitudes of type $A_{jk}(t)$ in \cite{p:2} in
terms of known special functions. The same assumptions were used in
\cite{p:0} (see \cite{p:0}, relations (37) -- (39) and (42) -- (47))
and will be used in this paper. Note that putting $g(m-m_{g}) \equiv
1$ in (\ref{A-SS}) leads to strictly exponential form of amplitudes
of type ${\cal A}_{SS}(t)$ as functions of time $t$. On the other
hand, keeping $g(m)$ in the assumed simplest physically admissible
form (\ref{g(m)}) results in the presence of additional
nonoscillatory terms in amplitudes of type ${\cal A}_{SS}(t), {\cal
A}_{LL}(t)$ etc. and thus in amplitudes $A_{jk}(t)$ as well (see
\cite{p:2,p:0}).

In \cite{p:0} the analytical formulae for $A_{jk}(t)$ obtained in
\cite{p:2} were used as the starting point to find analytical
expressions for matrix elements of the effective Hamiltonian for
this model for $t=\tau_{L}$ and then to obtain a numerical value for
the possible consequence of the Khalfin's Theorem analyzed in
\cite{p:1}. It is found there that, contrary to the standard LOY
result (\ref{j1-14}), the diagonal matrix elements of the exact
effective Hamiltonian for neutral meson complex cannot be equal if
CPT symmetry holds but CP symmetry is violated. We found in
\cite{p:0} that
\begin{equation}
\Re (h_{11}(t\sim \tau_{L})-h_{22}(t\sim \tau_{L}))\simeq
-4.771\times 10^{-18} MeV \label{j4-1},
\end{equation}
\begin{equation}
\Im (h_{11}(t\sim \tau_{L})-h_{22}(t\sim \tau_{L}))\simeq
7.283\times 10^{-16} MeV \label{j4-2}
\end{equation}
and
\begin{equation}
\frac{|\Re (h_{11}(t\sim \tau_{L})-h_{22}(t\sim
\tau_{L}))|}{m_{average}}\equiv
\frac{m_{K^{0}}-m_{\bar{K^{0}}}}{m_{average}}\sim 10^{-21}
\label{j4-3},
\end{equation}
where $h_{jk}(t)=\langle j | H_{\parallel}(t) | k \rangle$ ($j, k=1,
2$) and $H_{\parallel}(t)$ is the effective Hamiltonian. These
results were obtained analytically for the considered model for the
neutral kaon system in the case when the total system is CPT --
invariant but CP -- non--invariant (equations (68), (69) and (70) in
\cite{p:0}). The estimations (\ref{j4-1}) -- (\ref{j4-3}) were
obtained by inserting (\ref{j1-73}) and related $m_{S}\simeq
m_{L}\simeq m_{average} = 497.648 MeV ,$ $\Delta m = 3.489 \times
10^{-12} MeV,$ $\tau_{S} = 0.8935 \times 10^{-10} s ,$ $\tau_{L}=
5.17 \times 10^{-8} s ,$ $\gamma_{L}=1.3 \times 10^{-14} MeV,$
$\gamma_{S}=7.4 \times 10^{-12} MeV $ in formulae of type
(\ref{j1-72}). In this paper we will use the same experimental data.
We will also use the same notations and definitions as in
\cite{p:0}:
\begin{equation}
\gamma_{S}\equiv \frac{\Gamma_{S}}{2}, \ \gamma_{L}\equiv
\frac{\Gamma_{L}}{2}, \  \Delta m\equiv m_{L}-m_{S}, \label{j1-74}
\end{equation}
and so on. Note that results (\ref{j4-1}) -- (\ref{j4-3}) agree with
the general result obtained in \cite{p:1}.

The detailed analysis of the matrix elements of the effective
Hamiltonian for the $K^{0}-\bar{K^{0}}$ system shows that the non --
zero difference between the diagonal matrix elements of the
effective Hamiltonian in the considered model is caused by the
nonzero contribution into $\rho_{jk}(m)$, (\ref{rho-jk}), coming
from expressions for  $\langle
K_{S}|e^{\textstyle{-itH}}|K_{L}\rangle$ and $\langle
K_{L}|e^{\textstyle{-itH}}|K_{S}\rangle$ and by the non-oscillatory
terms in the formulae for the amplitudes of type (\ref{j1-37ab}) for
transitions: $K^{0}\longleftrightarrow K^{0}$,
$\bar{K^{0}}\longleftrightarrow \bar{K^{0}}$
$K^{0}\longleftrightarrow \bar{K^{0}}$. It is not difficult to
verify that neglecting the mentioned nonzero contribution and
dropping all these non-oscillatory terms leads to the zero
difference of the diagonal matrix elements of the effective
Hamiltonian in the considered case. This is because, in fact,
dropping these non--oscillatory terms is equivalent to replacing in
(\ref{j1-37ab}) densities $\rho_{jk}(m)$ defined by (\ref{rho-jk})
-- (\ref{g(m)}) with densities defined by the new function
$g_{WW}(m)$ instead of $g(m)$ given by (\ref{g(m)}) such that
$g_{WW}(m) = 1$ for all $ - \infty \leq m \leq + \infty$. Thus, the
integrals, e.g. in formulae of type (\ref{A-SS}), are taken between
the limits $m= - \infty$ and $m = + \infty$ with densities of type
(\ref{rho-SS}) having the Breit--Wigner form (and not truncated for
$m < m_{g} =0$) which leads to strictly exponential form of, e.g.
$|{\cal A}_{SS}(t)|^{2}$ and the like. The effective Hamiltonian
$H_{\parallel}(t)$ obtained in such a case is the LOY effective
Hamiltonian, $H_{LOY}$.

In this paper we continue searching for the properties of the model
analyzed in \cite{p:0}. The aim is to show how the difference of the
diagonal matrix elements $(h_{11}(t)-h_{22}(t))$ discussed in
\cite{p:0} and some other parameters describing neutral kaons
(including $r(t)$, (\ref{khalfin})) change in time $t$ in the case
of preserved CPT and violated CP symmetries. The paper is organized
as follows. In Section 2 we collect the formulae for the matrix
elements of the effective Hamiltonian necessary for further
analysis. Section 3 contains a numerical verification of the
Khalfin's Theorem (\ref{khalfin}). We show there how this Theorem
"acts". In Section 4 the value of the difference of the diagonal
elements $(h_{11}(t)-h_{22}(t))$ is shown as calculated at
$t=\tau_{L}$ with the use of the Mathematica. Also, in this Section
the time dependence of the real and imaginary parts of the diagonal
matrix elements of the effective Hamiltonian $(h_{11}(t)-h_{22}(t))$
in graphical form is given. In Section 5 the eigenvalues of the
effective Hamiltonian $\mu_{L}(t), \mu_{S}(t)$ are calculated and
the parameters of the violation of the CP symmetry
$\varepsilon_{L}(t), \varepsilon_{S}(t)$ are estimated. We also show
graphically the time dependence of all the calculated quantities
there. In Section 6 we check the correctness of our results by
verifying  the relation
$(\mu_{L}(t)+\mu_{S}(t)=h_{11}(t)+h_{22}(t))$ known from the
literature. Section 7 contains a discussion of the results obtained
in Sections 3 -- 6 and some remarks concerning the experiments with
neutral kaons.

\section{Matrix elements of the effective Hamiltonian}

General conclusions concerning properties of matrix elements of the
effective Hamiltonian $H_{\parallel}(t)$ can be drawn using the
following identity \cite{p:1}
\begin{equation}
H_{\parallel}(t)\equiv i\frac{\partial \textbf{A}(t)}{\partial
t}[\textbf{A}(t)]^{-1}, \label{j1-37}
\end{equation}
where all matrix elements $A_{jk}(t),$ ($j,k=1,2$) of the matrix
$\textbf{A}(t)$ can be calculated, e.g.  by means of
(\ref{j1-37ab}).

Using (\ref{j1-37}), one can calculate all the matrix elements of
the effective Hamiltonian  $H_{\parallel}$ which may now be written
as
\begin{eqnarray}
h_{11}(t)= \frac{i}{det \textbf{A}(t)}\cdot \biggl(\frac{\partial
A_{11}(t)}{\partial t} \cdot A_{22}(t)-\frac{\partial
A_{12}(t)}{\partial t}\cdot A_{21}(t)\biggl), \label{jz-1}
\end{eqnarray}

\begin{eqnarray}
h_{12}(t)= \frac{i}{det \textbf{A}(t)} \cdot \biggl(-\frac{\partial
A_{11}(t)}{\partial t}\cdot A_{12}(t)+\frac{\partial
A_{12}(t)}{\partial t}\cdot A_{11}(t)\biggl), \label{jz-2}
\end{eqnarray}

\begin{eqnarray}
h_{21}(t)=\frac{i}{det \textbf{A}(t)}\cdot \biggl(\frac{\partial
A_{21}(t)}{\partial t}\cdot A_{22}(t)-\frac{\partial
A_{22}(t)}{\partial t}\cdot A_{21}(t)\biggl), \label{jz-3}
\end{eqnarray}

\begin{eqnarray}
h_{22}(t)=\frac{i}{det \textbf{A}(t)} \cdot \biggl(-\frac{\partial
A_{21}(t)}{\partial t}\cdot A_{12}(t)+\frac{\partial
A_{22}(t)}{\partial t}\cdot A_{11}(t)\biggl), \label{jz-4}
\end{eqnarray}
where
\begin{eqnarray}
det \textbf{A}(t)= A_{11}(t) \cdot A_{22}(t)-A_{12}(t) \cdot
A_{21}(t). \label{jz-5a}
\end{eqnarray}

So, having these relations and inserting analytical expressions for
$A_{jk}(t)$, (\ref{j1-37ab}), calculated in \cite{p:0} within the
assumptions (\ref{j1-3}) and $[CP, H] \neq 0,$ one obtains all
matrix elements $h_{jk}(t)$ for the model considered. Next, such
obtained analytical formulae for $h_{jk}(t)$ can be used for
numerical calculations of some parameters characterizing neutral
kaons for instants of time changing in given time intervals.

\section{Numerical examination \\of the Khalfin's Theorem}

It seems to be interesting to verify how the Khalfin's Theorem
(\ref{khalfin}) acts in the system on neutral mesons. To see this,
we can use amplitudes $A_{12}(t)$ and $A_{21}(t)$ calculated
within the model considered in \cite{p:0} in the case of conserved
CPT and violated CP symmetries.  It is not difficult to calculate
the modulus of the ratio $\frac{A_{12}(t)}{A_{21}(t)}$ using
numerical methods. The results of such calculations are presented
below in Fig.\ref{TK3} and Fig.\ref{TK5}. There are $y(x) =
|\frac{A_{12}(x)}{A_{21}(x)}|$ and
$x=\frac{\gamma_{L}}{\hbar}\cdot
t$ in these figures. \\

\begin{figure}[H]
\begin{center}
\includegraphics[width=110mm]{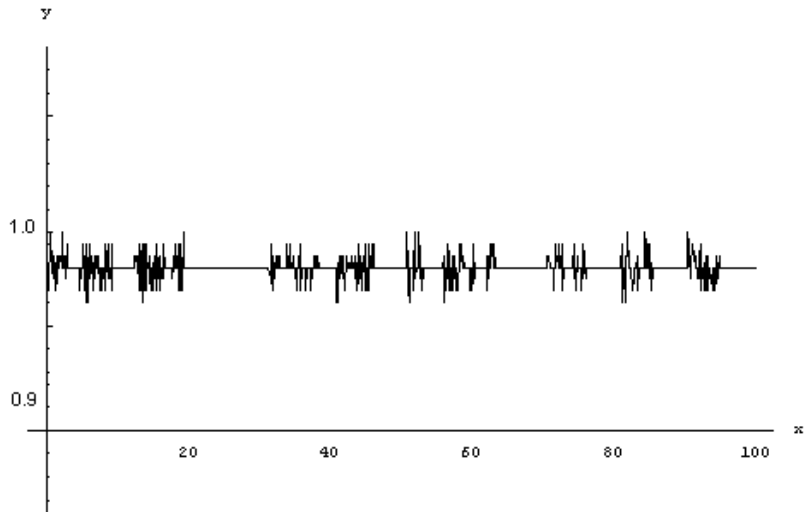}
\caption{The time dependence of the absolute value of
$y(x)=|r(t)|\equiv |\frac{A_{12}(t)}{A_{21}(t)}|$ in $x\in
(0.01,103).$ Here and in all other Figures:
$x=\frac{\gamma_{L}}{\hbar}\cdot t.$}\label{TK3}
\end{center}
\end{figure}

\begin{figure}[H]
\begin{center}
\includegraphics[width=110mm]{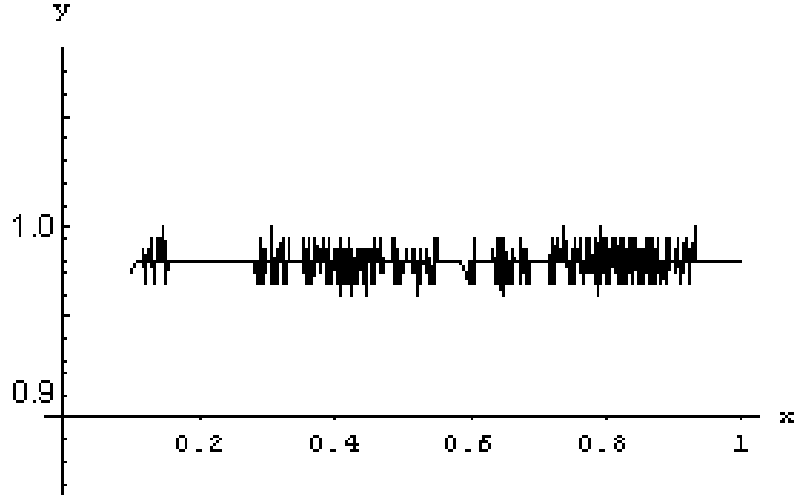}
\caption{The time dependence of the absolute value of
$y(x)=|r(t)|\equiv |\frac{A_{12}(t)}{A_{21}(t)}|$ in $x\in
(0.1,1).$}\label{TK5}
\end{center}
\end{figure}

These figures show that if one is able to measure the modulus of the
ratio $\frac{A_{12}(t)}{A_{21}(t)}$ only up to the accuracy of the
order $10^{-15}$ then one sees this ratio as a constant function of
time: for $x\in (0.01,103)$ we find that
\begin{eqnarray}
y_{max}(x)-y_{min}(x)=3.33067 \times 10^{-16}, \label{tk3a}
\end{eqnarray}
where
\begin{eqnarray}
y_{max}(x)&=&|r(t)|_{max}, \nonumber \\
y_{min}(x)&=&|r(t)|_{min}. \label{tk3c}
\end{eqnarray}

\section{ The difference of the diagonal matrix elements
 $(h_{11}(t)-h_{22}(t))$}

Assuming that the CPT symmetry is conserved in the system under
considerations ($[CPT, H]=0$) and using the necessary relations from
\cite{p:0, p:2, p:1} one finds the general form of the difference of
the diagonal matrix elements of the effective Hamiltonian. It has
the following form

\begin{eqnarray}
h_{11}(t)-h_{22}(t)=\frac{X(t)}{det \textbf{A}(t)}, \label{j1-79}
\end{eqnarray}
where
\begin{eqnarray}
X(t)= i\cdot \left(\frac{\partial A_{21}(t)}{\partial t}\cdot
A_{12}(t)-\frac{\partial A_{12}(t)}{\partial t}\cdot
A_{21}(t)\right) \label{j1-79a}
\end{eqnarray}
and $det \textbf{A}(t)$ is defined in (\ref{jz-5a}).

Our analytical result in the one-pole approximation obtained in
\cite{p:0} for $t=\tau_{L}$ can be written as
\begin{eqnarray}
h_{11}( \tau_{L})-h_{22}(\tau_{L}) \simeq(-4.771\times 10^{-18}+i
\cdot 7.283\times 10^{-16}) MeV. \label{jz-6}
\end{eqnarray}

The numerical result for $t=\tau_{L}$ in the one-pole approximation
obtained using the Mathematica has the following form
\begin{eqnarray}
h_{11}(\tau_{L})-h_{22}( \tau_{L})\simeq(-7.129\times 10^{-17}+i
\cdot 1.986\times 10^{-13}) MeV . \label{jz-7}
\end{eqnarray}

It is seen, that the difference between (\ref{jz-6}) and
(\ref{jz-7}) is small and it may be attributed to finite accuracy of
numerical calculations performed by Mathematica. No approximations
have been used in the analytical calculations.

Putting  $A_{jk}(t)$ ($j,k=1,2$) given by (\ref{j1-37ab}) into
(\ref{j1-79}) and using the energy density $\rho_{jk}(m)$ found in
\cite{p:0}, the difference $(h_{11}(t)-h_{22}(t))$ can be calculated
as the function ot time $t$. The results of our calculations are
presented in a graphical form. The figures below show the time
dependence of the real and imaginary parts of the diagonal matrix
elements of the effective Hamiltonian $(h_{11}(t)-h_{22}(t)).$
\pagebreak

\begin{figure}[H]
\begin{center}
\includegraphics[width=110mm]{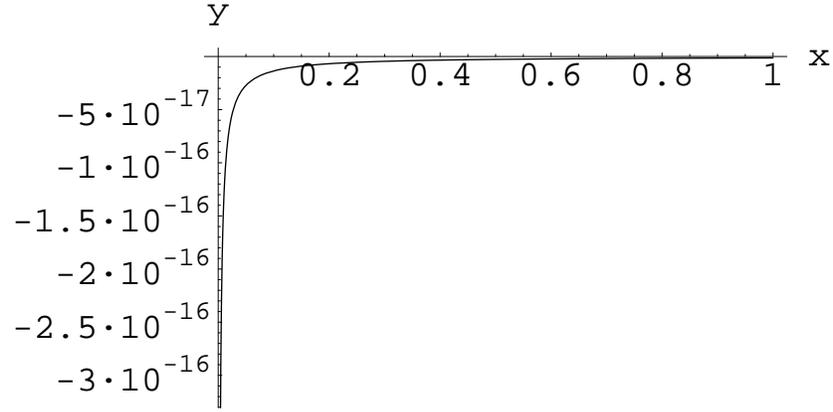}
\caption{The time dependence of the real part of the diagonal matrix
elements of the effective Hamiltonian $y=Re(h_{11}(x)-h_{22}(x))$ in
the range $x\in (0.001, 1).$}\label{wykres1a}
\end{center}
\end{figure}

Note that in  Fig.\ref{wykres1a} we have
$Re(h_{11}(t)-h_{22}(t))=Re(h_{11}(\tau_{L})-h_{22}(\tau_{L}))$ for
$x=1.$\\

\begin{figure}[H]
\begin{center}
\includegraphics[width=110mm]{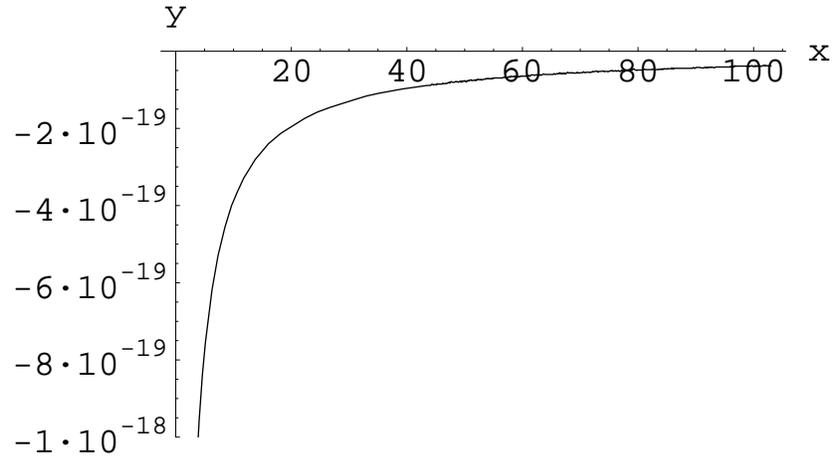}
\caption{The time dependence of the real part of the diagonal matrix
elements of the effective Hamiltonian $y=Re(h_{11}(x)-h_{22}(x))$ in
the interval $x\in (1, 103).$}\label{wykres1b}
\end{center}
\end{figure}

\begin{figure}[H]
\begin{center}
\includegraphics[width=110mm]{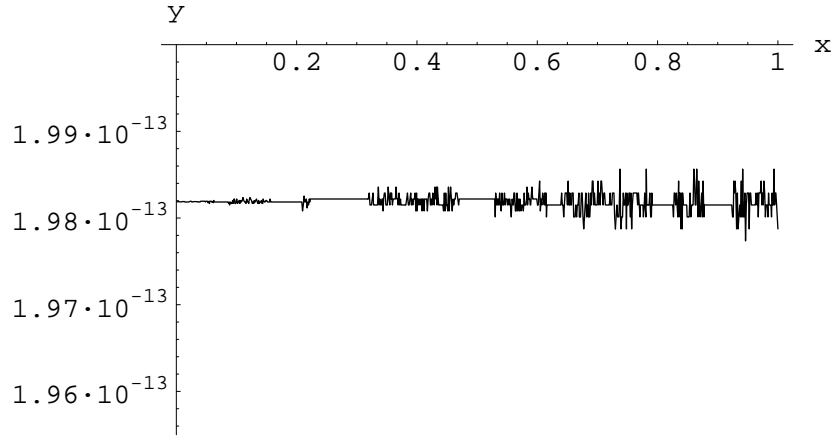}
\caption{The time dependence of the imaginary part of the diagonal
matrix elements of the effective Hamiltonian
$y=Im(h_{11}(x)-h_{22}(x))$ in the interval $x\in (0.001,
1).$}\label{wykres2a}
\end{center}
\end{figure}

\begin{figure}[H]
\begin{center}
\includegraphics[width=110mm]{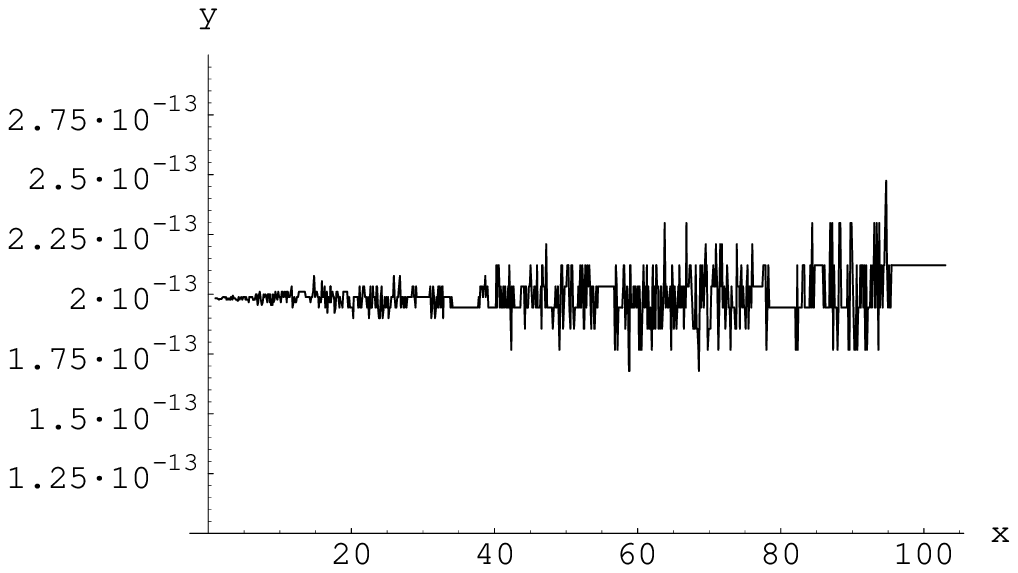}
\caption{The time dependence of the imaginary part of the diagonal
matrix elements of the effective Hamiltonian
$y=Im(h_{11}(x)-h_{22}(x))$ in the interval $x\in (1,
103).$}\label{wykres2b}
\end{center}
\end{figure}

At this point it should be explained that a more accurate analysis
of the results of the calculations which lead to Fig.\ref{wykres1a}
and Fig.\ref{wykres1b} and the use of a larger scale show that the
obtained curves are not so smooth as can be seen in
Fig.\ref{wykres2a} and Fig.\ref{wykres2b} but they are similar to
curves in Fig.\ref{wykres2a} and Fig.\ref{wykres2b}.

\section{Calculation of $\mu_{L}(t), \mu_{S}(t)$
and $\varepsilon_{L}(t), \varepsilon_{S}(t)$}

The eigenvalues of the effective Hamiltonian $\mu_{L}(t),
\mu_{S}(t)$ can be written as \cite{j2:7}
\begin{eqnarray}
\mu_{L}(t)=h_{o}(t)-h(t) , \label{jz-23}
\end{eqnarray}
\begin{eqnarray}
\mu_{S}(t)=h_{o}(t)+h(t) , \label{jz-24}
\end{eqnarray}
where
\begin{eqnarray}
h_{o}(t)=\frac{1}{2} \cdot (h_{11}(t)+h_{22}(t)) , \label{jz-25}
\end{eqnarray}
\begin{eqnarray}
h(t)=\sqrt{h_{z}^{2}(t)+h_{12}(t)\cdot h_{21}(t)}  \label{jz-26}
\end{eqnarray}
and
\begin{eqnarray}
h_{z}(t)=\frac{1}{2} \cdot (h_{11}(t)-h_{22}(t)).  \label{jz-27}
\end{eqnarray}

From (\ref{jz-23}) and (\ref{jz-24}) we have
\begin{eqnarray}
\mu_{S}(t) + \mu_{L}(t)=h_{11}(t)+h_{22}(t)\equiv
Tr(H_{\parallel}(t)). \label{jz-23a}
\end{eqnarray}
Relation (\ref{jz-23a}) does not depend on any approximations and it
is always true for every $(2 \times 2) $ matrix. Inserting
(\ref{jz-1}) - (\ref{jz-4}) into (\ref{jz-23}) and (\ref{jz-5a}) and
then using (\ref{j1-37ab}) and performing all integrations of type
(\ref{j1-37ab}) one can obtain for $t=\tau_{L}$
\begin{eqnarray}
\mu_{L}(\tau_{L})\simeq(497.648- i \cdot 4.458\times 10^{-13}) MeV,
\label{jz-8}
\end{eqnarray}
and
\begin{eqnarray}
\mu_{S}(\tau_{L})\simeq(497.648-i \cdot 2.471\times 10^{-13}) MeV .
\label{jz-9}
\end{eqnarray}
The general formula for $\mu_{L(S)}(t)$ can also be written as
follows
\begin{eqnarray}
\mu_{L(S)}(t)=m_{L(S)}(t)-\frac{i}{2} \ \ \gamma_{L(S)}(t).
\label{jz-9a}
\end{eqnarray}

The results of our calculations for the real part and imaginary part
in (\ref{jz-8}) and (\ref{jz-9}) are rounded to the third decimal
place. It should be noted that the real part in (\ref{jz-8}) and the
real part in (\ref{jz-9}) differ in the fourteenth decimal place.
The above mentioned result corresponds with the fact, that
$m_{S}\neq m_{L}$ and there is $|m_{L}-m_{S}| \sim |\gamma_{S}|,$
\cite{p:14, p:31}.

The time dependence of $\mu_{L}(t)$ and $\mu_{L}(t)$ is given below.

\begin{figure}[H]
\begin{center}
\includegraphics[width=110mm]{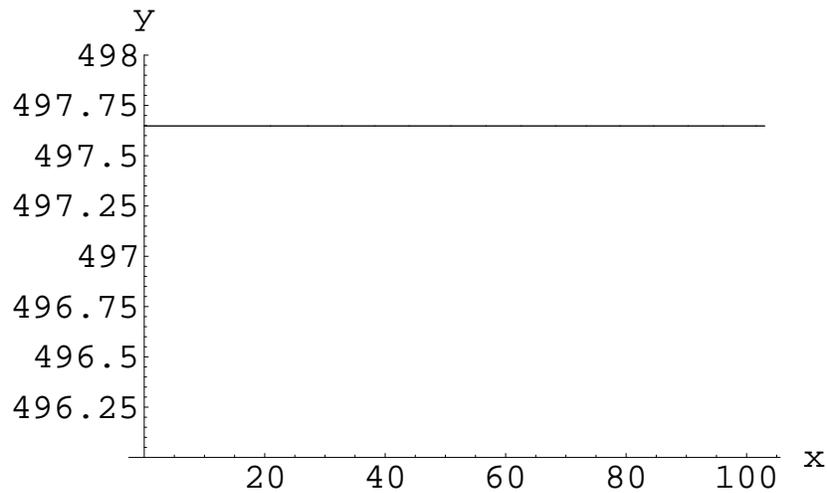}
\caption{The time dependence of the real part of $\mu_{L}(x):$
$y=Re(\mu_{L}(x))$ in the interval $x\in (0.001,
103).$}\label{wykres3}
\end{center}
\end{figure}

\begin{figure}[H]
\begin{center}
\includegraphics[width=110mm]{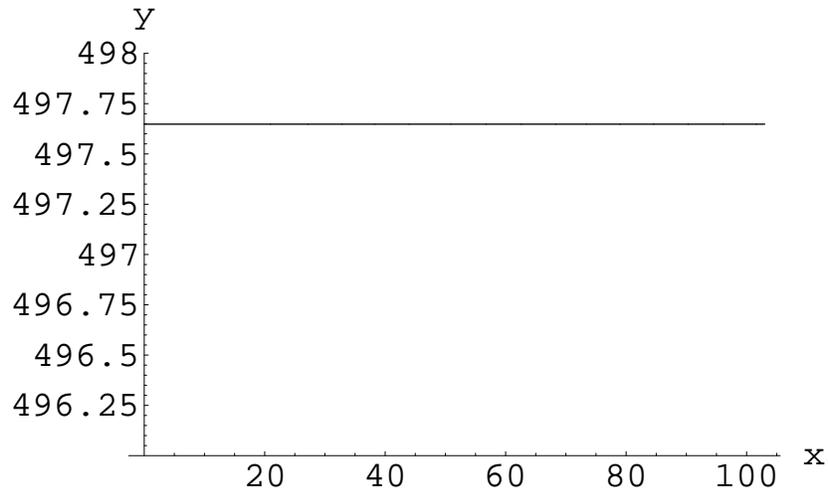}
\caption{The time dependence of the real part of $\mu_{S}(x):$
$y=Re(\mu_{S}(x))$ in the interval $x\in (0.001,
103).$}\label{wykres4}
\end{center}
\end{figure}

Expansion of scale in Fig.\ref{wykres3} and Fig.\ref{wykres4} shows
that continuous fluctuations with amplitudes of the order of
$10^{-14}$ appear.

\begin{figure}[H]
\begin{center}
\includegraphics[width=110mm]{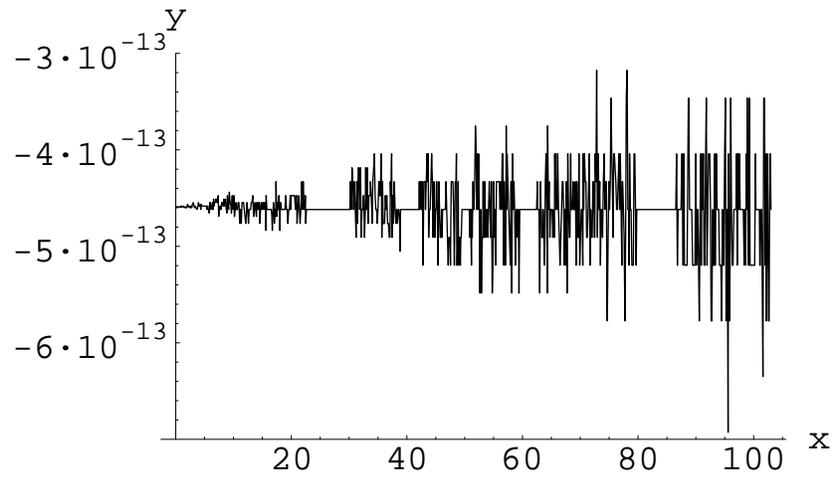}
\caption{The time dependence of the imaginary part of $\mu_{L}(x):$
$y=Im(\mu_{L}(x))$ in the interval $x\in (0.001,
103).$}\label{wykres5}
\end{center}
\end{figure}

\begin{figure}[H]
\begin{center}
\includegraphics[width=110mm]{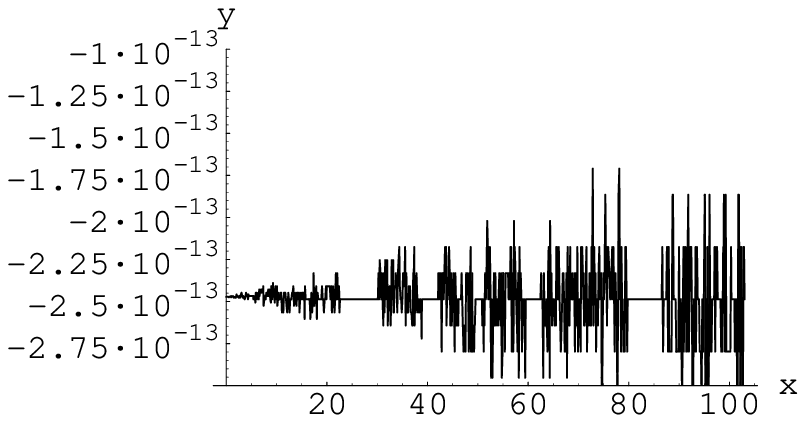}
\caption{The time dependence of the imaginary part of $\mu_{S}(x):$
$y=Im(\mu_{S}(x))$ in the interval $x\in (0.001,
103).$}\label{wykres6}
\end{center}
\end{figure}

We have the following formulae (see, e.g. \cite{j2:7})
\begin{eqnarray}
\varepsilon_{L}(t)=-
\frac{h_{21}(t)-h_{22}(t)+\mu_{L}(t)}{h_{21}(t)+h_{22}(t)-\mu_{L}(t)}
\label{jz-16}
\end{eqnarray}
\begin{eqnarray}
\varepsilon_{S}(t)=-
\frac{h_{21}(t)+h_{22}(t)-\mu_{S}(t)}{h_{21}(t)-h_{22}(t)+\mu_{S}(t)}
\label{jz-17}
\end{eqnarray}
and we get for $t=\tau_{L}$
\begin{eqnarray}
\varepsilon_{L}(\tau_{L})\simeq -1.0000000000184743'+ i \cdot 0.0'
\label{jz-18}
\end{eqnarray}
and
\begin{eqnarray}
\varepsilon_{S}(\tau_{L})\simeq 1.0000157759810688'- i \cdot 0.0'
\label{jz-19}
\end{eqnarray}

The time dependence of $\varepsilon_{L}(t)$ and $\varepsilon_{S}(t)$
are presented below.

\begin{figure}[H]
\begin{center}
\includegraphics[width=110mm]{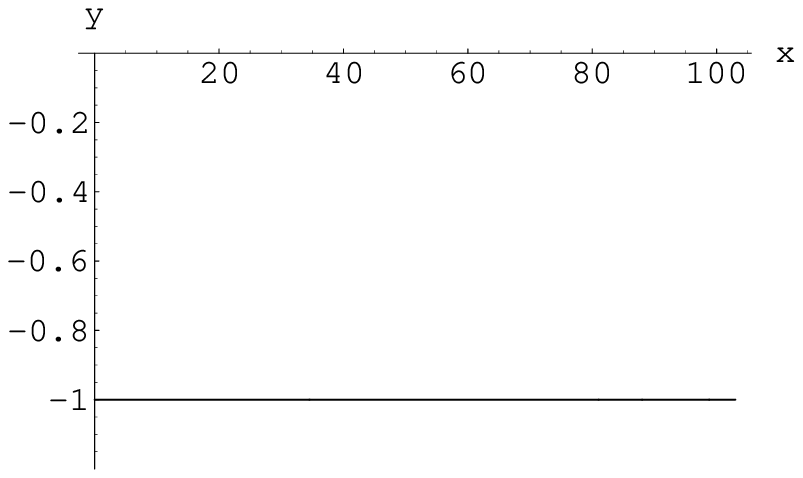}
\caption{The time dependence of the real part of
$\varepsilon_{L}(x):$ $y=Re(\varepsilon_{L}(x))$ in the interval
$x\in (0.001, 103).$}\label{wykres7}
\end{center}
\end{figure}

\begin{figure}[H]
\begin{center}
\includegraphics[width=110mm]{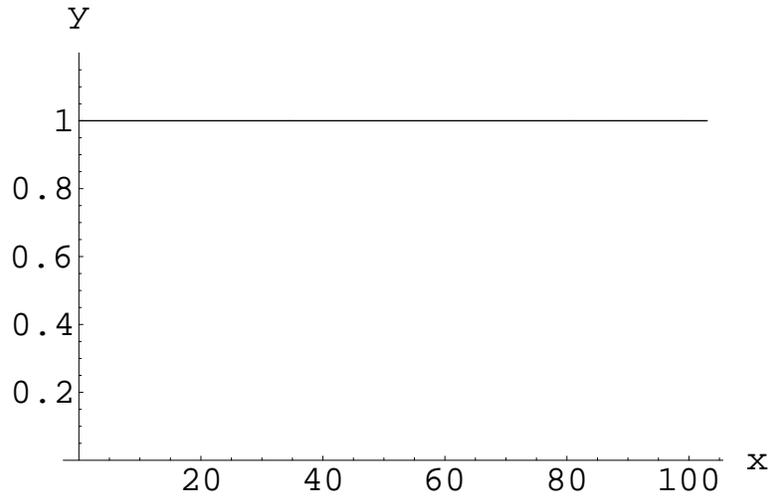}
\caption{The time dependence of the real part of
$\varepsilon_{S}(x):$ $y=Re(\varepsilon_{S}(x))$ in the interval
$x\in (0.001, 103).$}\label{wykres8}
\end{center}
\end{figure}

Expansion of scale on Fig.\ref{wykres7} and Fig.\ref{wykres8} shows,
that continuous fluctuations with amplitudes of the order of
$10^{-12}$ appear here.

\begin{figure}[H]
\begin{center}
\includegraphics[width=110mm]{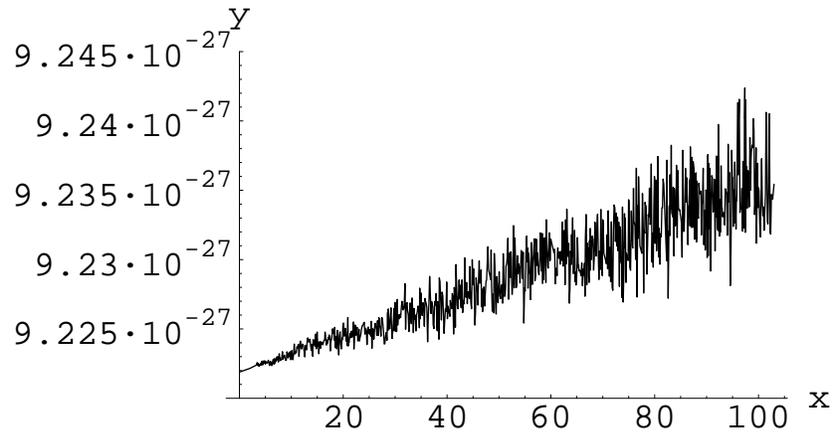}
\caption{The time dependence of the imaginary part of
$\varepsilon_{L}(x):$ $y=Im(\varepsilon_{L}(x))$ in the interval
$x\in (0.001, 103)$}\label{wykres9}
\end{center}
\end{figure}

\begin{figure}[H]
\begin{center}
\includegraphics[width=110mm]{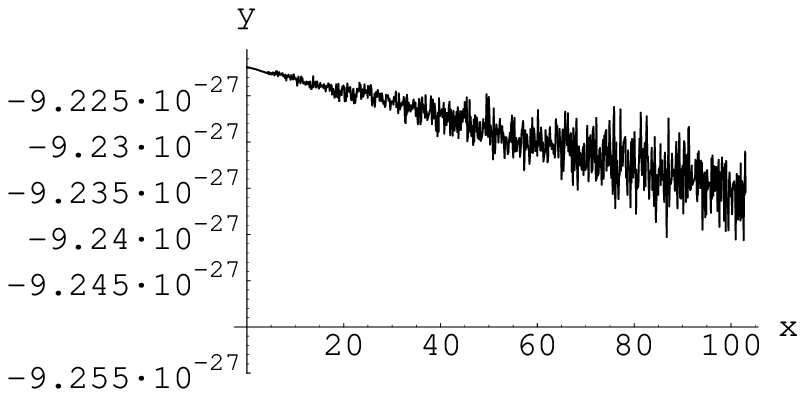}
\caption{The time dependence of the imaginary part of
$\varepsilon_{S}(x):$ $y=Im(\varepsilon_{S}(x))$ in the interval
$x\in (0.001, 103)$}\label{wykres10}
\end{center}
\end{figure}

From the formula
\begin{eqnarray}
\varepsilon(t)=\frac{1}{2}\cdot
(\varepsilon_{L}(t)+\varepsilon_{S}(t)) \label{jz-20}
\end{eqnarray}
we have for $t=\tau_{L}$
\begin{eqnarray}
\varepsilon(\tau_{L})\simeq7.888 \times 10^{-6}- i \cdot 0.0' .
\label{jz-21}
\end{eqnarray}

The absolute value of $\varepsilon(\tau_{L})$
\begin{eqnarray}
|\varepsilon(\tau_{L}) |\simeq7.888 \times 10^{-6} . \label{jz-22}
\end{eqnarray}

The figures below present the time dependence of the absolute value
of $\varepsilon(t).$

\begin{figure}[H]
\begin{center}
\includegraphics[width=110mm]{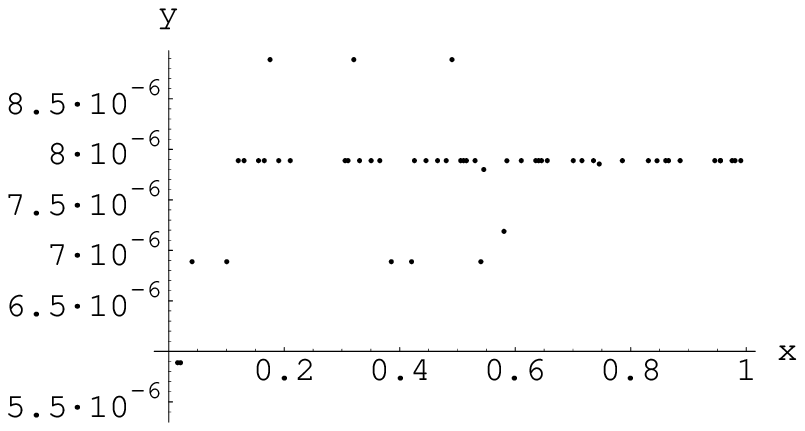}
\caption{The time dependence of the absolute value of
$\varepsilon(t)$: $y=|\varepsilon(t)|$ in the interval $x\in (0.001,
103).$}\label{wykres11}
\end{center}
\end{figure}

\begin{figure}[H]
\begin{center}
\includegraphics[width=110mm]{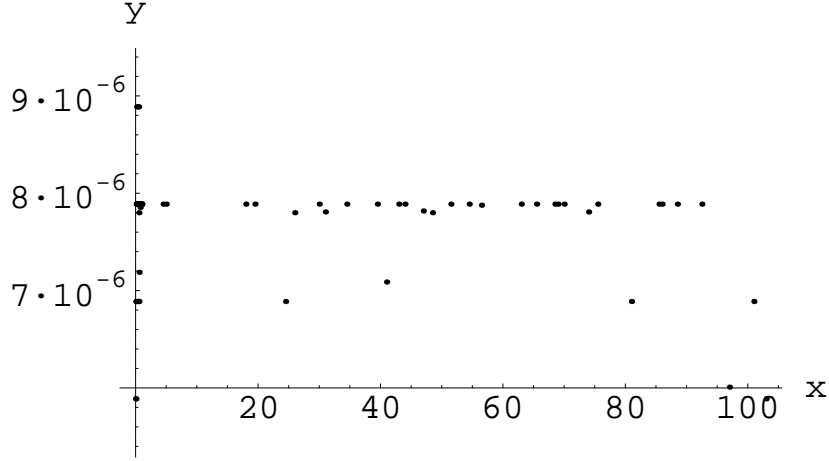}
\caption{The time dependence of the absolute value of
$\varepsilon(t)$: $y=|\varepsilon(t)|$ in the interval $x\in (1,
103).$ }\label{wykres12}
\end{center}
\end{figure}

\section{Verification of the relation
$\mu_{L}(t)+\mu_{S}(t)=h_{11}(t)+h_{22}(t)$}

All results in this Section have been rounded to the decimal third
place.

In accordance with formulae ((\ref{jz-8}), (\ref{jz-9})) for
$t=\tau_{L}$ we have
\begin{eqnarray}
\mu_{L}(\tau_{L})\simeq(497.648- i \cdot 4.458\times 10^{-13}) MeV,
\nonumber
\end{eqnarray}
\begin{eqnarray}
\mu_{S}(\tau_{L})\simeq(497.648-i \cdot 2.471\times 10^{-13}) MeV
\nonumber
\end{eqnarray}
and the corresponding matrix elements of the effective Hamiltonian
(formulae (\ref{jz-1})-(\ref{jz-4})) can be written for $t=\tau_{L}$
as
\begin{eqnarray}
h_{11}(\tau_{L})\simeq(497.648-i \cdot 2.471\times 10^{-13}) MeV ,
\label{jz-12}
\end{eqnarray}
\begin{eqnarray}
h_{12}(\tau_{L})\simeq(1.787\times 10^{-23}-i \cdot 6.401\times
10^{-24}) MeV ,\label{jz-13}
\end{eqnarray}
\begin{eqnarray}
h_{21}(\tau_{L})\simeq(-1.799\times 10^{-23}-i \cdot 5.127\times
10^{-24}) MeV ,\label{jz-14}
\end{eqnarray}
\begin{eqnarray}
h_{22}(\tau_{L})\simeq(497.648-i \cdot 4.458\times 10^{-13}) MeV
.\label{jz-15}
\end{eqnarray}
For $t=\tau_{L}$ we get
\begin{eqnarray}
\mu_{L}(\tau_{L})+\mu_{S}(\tau_{L})=h_{11}(\tau_{L})+h_{22}(\tau_{L})\simeq(995.296-i
\cdot 6.929\times 10^{-13}) MeV. \label{jz-11}
\end{eqnarray}
Relation (\ref{jz-23a}) is also fulfilled at $t=\tau_{S}$
\begin{eqnarray}
\mu_{L}(\tau_{S})+\mu_{S}(\tau_{S})=h_{11}(\tau_{S})+h_{22}(\tau_{S})\simeq
(995.296 - i \cdot 9.623\times 10^{-14}) MeV. \label{jz-15a}
\end{eqnarray}
Comparing (\ref{jz-11}) with (\ref{jz-15a}) we can see that
\begin{eqnarray}
\mu_{L}(\tau_{L})+\mu_{S}(\tau_{L})\neq\mu_{L}(\tau_{S})+\mu_{S}(\tau_{S}).
\label{jz-15f1}
\end{eqnarray}
It is interesting to notice that
\begin{eqnarray}
\mu_{L}(\tau_{L})+\mu_{S}(\tau_{S})\simeq (995.296 - i \cdot
5.234\times 10^{-13}) MeV ,  \label{jz-15ab}
\end{eqnarray}
\begin{eqnarray}
\mu_{L}(\tau_{S})+\mu_{S}(\tau_{L})\simeq (995.296 - i \cdot
2.658\times 10^{-13}) MeV ,  \label{jz-15c}
\end{eqnarray}
\begin{eqnarray}
h_{11}(\tau_{S})+h_{22}(\tau_{L})\simeq (995.296 + i \cdot
4.644\times 10^{-13}) MeV ,  \label{jz-15d}
\end{eqnarray}
\begin{eqnarray}
h_{11}(\tau_{L})+h_{22}(\tau_{S})\simeq (995.296 + i \cdot
3.247\times 10^{-13}) MeV ,  \label{jz-15ae}
\end{eqnarray}
>From our calculations it follows that the real parts of formulae
(\ref{jz-15ab}) -- (\ref{jz-15ae}) differ in the twelfth or
thirteenth decimal place. This means that
\begin{eqnarray}
\mu_{L}(\tau_{S})+\mu_{S}(\tau_{L})\neq
\mu_{L}(\tau_{S})+\mu_{S}(\tau_{S}), \label{jz-15f}
\end{eqnarray}
\begin{eqnarray}
\mu_{L}(\tau_{S})+\mu_{S}(\tau_{L})\neq
\mu_{L}(\tau_{L})+\mu_{S}(\tau_{L}), \label{jz-15g}
\end{eqnarray}
\begin{eqnarray}
\mu_{L}(\tau_{S})+\mu_{S}(\tau_{L})\neq
\mu_{L}(\tau_{L})+\mu_{S}(\tau_{S}), \label{jz-15h}
\end{eqnarray}
\begin{eqnarray}
\mu_{L}(\tau_{S})+\mu_{S}(\tau_{L})\neq
h_{11}(\tau_{S})+h_{22}(\tau_{S}), \label{jz-15i}
\end{eqnarray}
\begin{eqnarray}
\mu_{L}(\tau_{L})+\mu_{S}(\tau_{S})\neq
h_{11}(\tau_{L})+h_{22}(\tau_{L}), \label{jz-15j}
\end{eqnarray}
\begin{eqnarray}
\mu_{L}(\tau_{L})+\mu_{S}(\tau_{S})\neq
h_{11}(\tau_{L})+h_{22}(\tau_{S}) \label{jz-15k}
\end{eqnarray}
and so on.

\section{Final remarks}

First,  as it was pointed out in \cite{p:2}, let us notice that in
the considered model some relations assumed there 
and allowing to perform integrations of type (\ref{j1-37ab}) are not
valid in the $K^{0}-\bar{K^{0}}$ system (see a comment between
formulae (5.10) and (5.11) of \cite{p:2}, Sec. 5). These relations
were also used in \cite{p:2} and \cite{p:0}. Next, a drawback of our
model is that at $t=0$ we obtain $h_{11}(t=0)=\infty $ and
$h_{22}(t=0)=\infty.$ However, this model allows to study all the
consequences of Khalfin's Theorem  and theorems considered in
\cite{p:1}. Bearing in mind the limitations of our model mentioned
above one should not expect that our calculations based on this
model will result in an exact reconstruction of all experimental
parameters characterizing the neutral K meson system.

Results presented in Sec. 3 show how Khalfin's Theorem works. We can
conclude that the effect of this Theorem should be visible in
experiments with the neutral kaon complex in which the modulus of
$r(t)$, (\ref{khalfin}), can be measured with the accuracy of order
$10^{-16}$ or better. On the other hand, these results are in
perfect agreement with the supposition formulated in
\cite{urbanowski-APP-B}. Experimental results give $|1 - |r(t)|\,|
\sim 10^{-3} =$ const with some limited accuracy (see, e.g.
\cite{p:14}). The explanation of this fact proposed in
\cite{urbanowski-APP-B} is based on the assumption that
\[
r(t) = r_{LOY} + d(t),
\]
where $r_{LOY}$ stands for $r(t)$ calculated within the LOY theory,
and $d(t)$ is assumed to be a function varying  in time $t$ such
that $ |d(t)| \,<\,10^{-11}$.

Results obtained in Section 4 suggest that the real part of the
difference $(h_{11}(t)-h_{22}(t))$ is different from zero for very
large times $t$: from $t \sim 0,1 \tau_{L}$ up to $ t \sim 100
\tau_{L}$. Moreover, after division by $m_{average},$ this
difference is only a little smaller than the corresponding
experimental value \cite{p:14}. The imaginary part of
$(h_{11}(t)-h_{22}(t))$ turned out to be different from zero as
well. However, this part oscillates about $2\cdot 10^{-13}$ MeV very
fast. Note that from the results contained in \cite{p:1} it follows
that these differences should differ from zero for all $t>0$. Within
the standard treatment of the neutral K system the measurement of
the difference of masses $(m_{K^{0}}-m_{\bar{K}^{0}})$ is considered
as the CPT invariance test. This interpretation of such tests is
based on the properties (\ref{j1-14}), (\ref{j1-15}) of the LOY
approach: Within the LOY theory CPT symmetry is conserved only if
$(m_{K^{0}}-m_{\bar{K}^{0}}) = 0 $.  The results obtained in Sec. 4
and in \cite{p:0,p:1,epjc-2007} show that such an interpretation of
this test is true only for the LOY approximation and beyond LOY
approximation properties (\ref{j1-14}), (\ref{j1-15}) do not occur.
It seems to be obvious that the description of neutral $K$ complex
using the more accurate formalism than the LOY approximation leads
to a more realistic description of such a complex. So, if within the
more accurate theory one obtains $(m_{K^{0}}-m_{\bar{K}^{0}}) \neq
0$ when CPT symmetry holds and CP symmetry is violated then one is
forced to conclude that such a property must be valid in the real
CPT invariant systems. Therefore, taking into account results
obtained in Sec. 4 (and in \cite{p:0,p:1,epjc-2007}) the conclusion
that the measurement of the mass difference,
$(m_{K^{0}}-m_{\bar{K}^{0}})$, should not be considered as CPT
invariance test seems to be correct.

Results in Section 5 show that the imaginary parts of parameters
 $\mu_{L}, \mu_{S}$ and $\varepsilon_{L},
\varepsilon_{S}$ vary in time too. We can say the same about their
real parts.  The oscillation amplitude is of the order of $10^{-13}$
for $Im\,(\mu_{L(S)})$ and it is smaller than $10^{-27}$ for
$Im\,(\varepsilon_{L(S)}).$ The real parts $Re\,(\mu_{L(S)})$ and
$Re\,(\varepsilon_{L(S)})$ oscillate in a similar way as their
imaginary parts and this is the reason why they cannot be shown in
our figures. The parameter $\varepsilon$, (\ref{jz-20}), is also a
quantity varying in time as we can see from the graphical results
(see Fig.\ref{wykres11} and Fig.\ref{wykres12}). The absolute value
of $|\varepsilon|$ oscillates around the value $8\cdot 10^{-6}$.

In Section 6, relation (\ref{jz-23a}) was investigated. We know,
from general considerations, that it has to be fulfilled for every
time $t$ irrespective of whether we consider an approximate model of
neutral K meson system or if we investigate the exactly solvable
model of this system. The considered mathematical results show that
the left side of equation (\ref{jz-23a}) is the same as its right
side for $t = \tau_{l}$, $t= \tau_{S}$. A similar mathematical
result was obtained for other times. From relation (\ref{jz-15f1})
it follows that the left side of equation (\ref{jz-23a}), as well as
its rights side, are not constant in the time.

Relations (\ref{jz-15ab}) -- (\ref{jz-15k}) show that equation
(\ref{jz-23a}) is no longer fulfilled when the separate components
of sums appearing in its left and right sides are taken at different
moments of time and then inserted into this equation. From this
observation and from results in Section 4, we can draw an important
conclusion concerning the methods of experimental data registering
and experimental data processing.

\begin{figure}[H]
\begin{center}
\includegraphics[width=110mm]{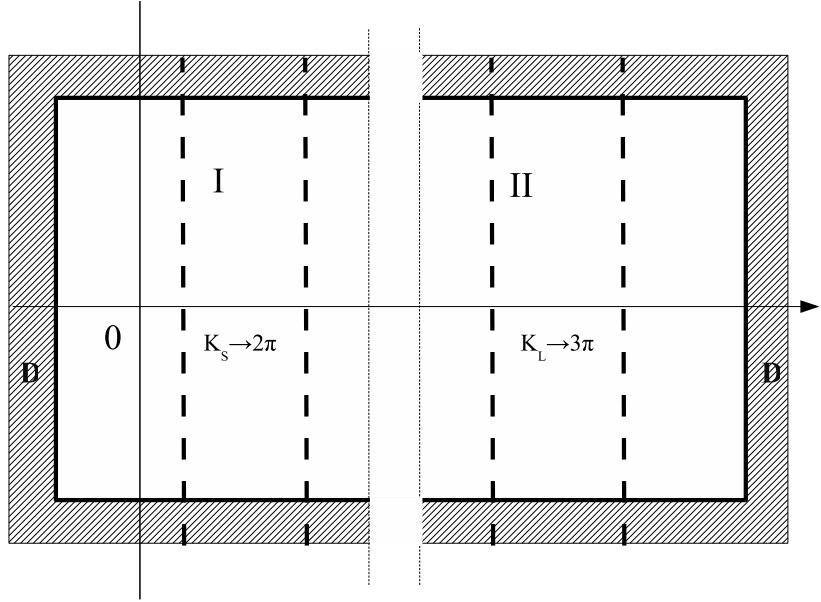}
\caption{A scheme of the experimental set for the experiment with
the neutral $K$ meson described in this section.}\label{rysunek}
\end{center}
\end{figure}

Let us note that an experimental system containing detectors which
register the neutral $K$ meson decay products can be schematically
presented as in Fig.\ref{rysunek}. This Figure presents a
longitudinal section of a cylindrical vacuum chamber. A $K$ meson
stream is fired into this chamber along a horizontal axis $l$ on the
left side. Detectors $D$ surrounding the chamber form its walls.
These detectors register the neutral $K$ meson decay products. In
the first region of this chamber (I), we observe a great amount of
the decay products of neutral kaons into two pions
($K_{S}\longrightarrow 2 \pi$). In the second region of this chamber
(II), we usually observe a great amount of the decay products of
neutral kaons into three pions ($K_{L}\longrightarrow 3 \pi$). We
can interpret the $l$ axis as a path in an uniform straight-line
motion of neutral kaons, whose decay products are registered by
detectors $D.$ We can write $l_{I}=v_{K_{S}} \cdot \tau_{S}$ in the
first region (I) and $l_{II}=v_{K_{L}} \cdot \tau_{L}$ in the second
region (II) (where $v_{K_{S(L)}}$ is the kaon $K_{S(L)}$ speed). One
can also obtain (for comparison): $l_{I}^{'}=c\cdot
\tau_{S}=0.026805 \ m$ and $l_{II}^{'}=c\cdot \tau_{L}=15.51 \ m$
($c=3 \times 10^{8} \ s$ - the speed of light in vacuum,
$\tau_{S}=0.8935 \times 10^{-10} \ s$ a $\tau_{L}=5.17 \times
10^{-8} \ s$).

Let us now return to the above mentioned conclusion. From
(\ref{jz-15ab}) -- (\ref{jz-15k}) and from results presented in
Section 5, it follows that only these parameters can correctly
reflect real properties of neutral $K$ system which are calculated
using only data obtained from a ring of detectors limited by
distances ($l,\;l + \Delta l$). Since $t \sim l$, the events are
registered between $t, \; t + \Delta t$ from the initial instant. Of
course, $\Delta l$ should be as small as possible. In other words,
one should not use the experimental data obtained from the
registration of the neutral $K$ meson decay products in the
calculations if these neutral $K$ meson decay products come from
different and distant parts of  the measurement set of the type
shown in Fig.\ref{rysunek}. For example, one should not use
calculations which were registered by the detector D in the region
where $t \sim \tau_{S}$ simultaneously with the data which were
registered by the detector D in the region where $t \sim \tau_{L}$.
One should not mix the data coming from different and distant parts
of the measurement. If this rule is not observed, it may turn out
that the obtained values $\mu_{L(S)}$ (obtained on the basis of
parameters measured in this way) will not satisfy the consistency
check given by (\ref{jz-23a}). Of course, in order to check the
consistency of the experimental results with (\ref{jz-23a}) the
experiment should be conducted in such a way that both sides of
(\ref{jz-23a}) can be found in independent measurements. Then one
will obtain independently each other the left side and the right
side of this equality.

\end{document}